\documentclass[conference]{IEEEtran}
\IEEEoverridecommandlockouts
\usepackage{cite}
\usepackage{amsmath,amssymb,amsfonts}
\usepackage{algorithmic}
\usepackage{graphicx}
\usepackage{textcomp}
\usepackage{xcolor}
\usepackage[nolist]{acronym}
\usepackage{paralist}
\usepackage{listings}
\usepackage{svg}
\usepackage{changepage} 
\usepackage{enumerate}
\def\BibTeX{{\rm B\kern-.05em{\sc i\kern-.025em b}\kern-.08em
    T\kern-.1667em\lower.7ex\hbox{E}\kern-.125emX}}

\makeatletter
\newcommand{\myitem}[1]{%
\item[#1]\protected@edef\@currentlabel{#1}%
}
\makeatother

\begin{document}

\makeatletter
\def\ps@IEEEtitlepagestyle{%
\def\@oddfoot{\parbox{\textwidth}{\footnotesize
Author's version of a paper accepted for publication in Proceedings of 2023 IEEE Belgrade PowerTech. 
\\
\textcopyright{} 2023 IEEE. 
Personal use of this material is permitted.  
Permission from IEEE must be obtained for all other uses, in any current or future media, including reprinting/republishing this material for advertising or promotional purposes, creating new collective works, for resale or redistribution to servers or lists, or reuse of any copyrighted component of this work in other works.\vspace{1.2em}}
}%
}
\makeatother

\begin{acronym}
\acro{sg}[SG]{smart grid}
\acroplural{sg}[SGs]{smart grids}
\acro{der}[DER]{distributed energy resource}
\acroplural{der}[DERs]{distributed energy resources}
\acro{ict}[ICT]{information and communication technology}
\acro{fdi}[FDI]{false data injection}
\acro{scada}[SCADA]{Supervisory Control and Data Acquisition}
\acro{mtu}[MTU]{Master Terminal Unit}
\acroplural{mtu}[MTUs]{Master Terminal Units}
\acro{hmi}[HMI]{Human Machine Interface}
\acro{plc}[PLC]{Programmable Logic Controller}
\acroplural{plc}[PLCs]{Programmable Logic Controllers}
\acro{ied}[IED]{Intelligent Electronic Device}
\acroplural{ied}[IEDs]{Intelligent Electronic Devices}
\acro{rtu}[RTU]{Remote Terminal Unit}
\acroplural{rtu}[RTUs]{Remote Terminal Units}
\acro{iec104}[IEC-104]{IEC 60870-5-104}
\acro{apdu}[APDU]{Application Protocol Data Unit}
\acro{apci}[APCI]{Application Protocol Control Information}
\acro{asdu}[ASDU]{Application Service Data Unit}
\acro{io}[IO]{information object}
\acroplural{io}[IOs]{information objects}
\acro{cot}[COT]{cause of transmission}
\acro{mitm}[MITM]{Man-in-the-Middle}
\acro{fdi}[FDI]{False Data Injection}
\acro{ids}[IDS]{intrusion detection system}
\acroplural{ids}[IDSs]{intrusion detection systems}
\acro{siem}[SIEM]{Security Information and Event Management}
\acro{mv}[MV]{medium voltage}
\acro{lv}[LV]{low voltage}
\acro{cdss}[CDSS]{controllable distribution secondary substation}
\acro{bss}[BSS]{battery storage system}
\acroplural{bss}[BSSs]{battery storage systems}
\acro{pv}[PV]{photovoltaic inverter}
\acro{mp}[MP]{measuring point}
\acroplural{mp}[MPs]{measuring points}
\acro{dsc}[DSC]{Dummy SCADA Client}
\acro{fcli}[FCLI]{Fronius CL inverter}
\acro{fipi}[FIPI]{Fronius IG+ inverter}
\acro{sii}[SII]{Sunny Island inverter}
\acro{tls}[TLS]{Transport Layer Security}
\acro{actcon}[ActCon]{Activation Confirmation}
\acro{actterm}[ActTerm]{Activation Termination}
\acro{rtt}[RTT]{Round Trip Time}
\acro{c2}[C2]{Command and Control}
\acro{dst}[DST]{Dempster Shafer Theory}
\acro{ec}[EC]{Event Correlator}
\acro{sc}[SC]{Strategy Correlator}
\acro{ioc}[IoC]{Indicator of Compromise}
\acroplural{ioc}[IoCs]{Indicators of Compromise}
\acro{ot}[OT]{Operational Technology}
\acro{soc}[SOC]{Security Operation Center}
\acro{pcap}[PCAP]{Packet Capture}
\acro{wcgan}[WCGAN]{Wasserstein Generative Adversarial Network}
\acro{dae}[DAE]{Deep Auto Encoder}
\acro{lstm}[LSTM]{Long Short-Term Memory}
\acro{mlp}[MLP]{MultiLayer Perceptron}
\acro{ics}[ICS]{Industrial Control System}
\acroplural{ics}[ICSs]{Industrial Control Systems}
\acro{pera}[PERA]{Purdue Enterprise Reference Architecture}
\acro{ews}[EWS]{Engineering Workstation}
\acroplural{ews}[EWSs]{Engineering Workstations}
\acro{hmi}[HMI]{Human-Machine Interface}
\acroplural{hmi}[HMIs]{Human-Machine Interfaces}
\acro{fp}[FP]{False Positive}
\acroplural{fp}[FPs]{False Positives}
\acro{fn}[FN]{False Negative}
\acroplural{fn}[FNs]{False Negatives}
\acro{tp}[TP]{True Positive}
\acroplural{tp}[TPs]{True Positives}
\acro{tn}[TN]{True Negative}
\acroplural{tn}[TNs]{True Negatives}
\acro{rdf}[RDF]{Resource Description Framework}
\acro{shacl}[SHACL]{Shapes Constraint Language}
\acro{pcap}[PCAP]{Packet Capture}
\acroplural{pcap}[PCAPs]{Packet Captures}
\acro{nist}[NIST]{National Institute of Standards and Technology}
\acro{ml}[ML]{Machine Learning}
\end{acronym}

\bstctlcite{IEEEexample:BSTcontrol}

\title{An Approach to Abstract Multi-stage Cyberattack Data Generation for ML-Based IDS in Smart Grids}

\author{
\IEEEauthorblockN{%
Ömer Sen\IEEEauthorrefmark{1},
Philipp Malskorn\IEEEauthorrefmark{1},
Simon Glomb\IEEEauthorrefmark{1},
Immanuel Hacker\IEEEauthorrefmark{1},
Martin Henze\IEEEauthorrefmark{2}\IEEEauthorrefmark{3},
Andreas Ulbig\IEEEauthorrefmark{1}
}

\IEEEauthorblockA{%
\IEEEauthorrefmark{1}\textit{IAEW, RWTH Aachen Univserity,} Aachen, Germany\\
Email: \{o.sen, i.hacker, a.ulbig\}@iaew.rwth-aachen.de, \{philipp.malskorn, simon.glomb\}@rwth-aachen.de}
\IEEEauthorblockA{%
\IEEEauthorrefmark{2}\textit{Security and Privacy in Industrial Cooperation, RWTH Aachen University,} Aachen, Germany\\
\IEEEauthorrefmark{3}\textit{Cyber Analysis \& Defense, Fraunhofer FKIE,} Wachtberg, Germany\\
Email: henze@cs.rwth-aachen.de}
}

\maketitle

\begin{abstract}
Power grids are becoming more digitized, resulting in new opportunities for the grid operation but also new challenges, such as new threats from the cyber-domain.
To address these challenges, cybersecurity solutions are being considered in the form of preventive, detective, and reactive measures.
Machine learning-based intrusion detection systems are used as part of detection efforts to detect and defend against cyberattacks.
However, training and testing data for these systems are often not available or suitable for use in machine learning models for detecting multi-stage cyberattacks in smart grids.
In this paper, we propose a method  to generate synthetic data using a graph-based approach for training machine learning models in smart grids.
We use an abstract form of multi-stage cyberattacks defined via graph formulations and simulate the propagation behavior of attacks in the network.
Within the selected scenarios, we observed promising results, but a larger number of scenarios need to be studied to draw a more informed conclusion about the suitability of synthesized data.
\end{abstract}


\section{Introduction} \label{sec:introduction}
The cyber-physical characteristic of the power grid, in particular the increasing penetration of \ac{ict}, addresses the issues of the distribution grid with respect to structural changes resulting from the integration of \ac{der}~\cite{vale2010distributed}.
In particular, it enables active grid operation at the distribution grid level and provides the backbone for the realization of \ac{sg} concepts~\cite{hossain2013smart,vandervelde2020methods}.
This circumstance provides not only new opportunities, but also new threats resulting from the increasing interconnectedness of systems and actors~\cite{yan2012survey}.
The new threat landscape consists not only of threats arising from cyber domain interdependencies, but also of new attack surfaces for cyberattacks that threaten the stability and reliability of the grid~\cite{mathas2020threat}.
To adequately address these new threats, defense and countermeasures must be integrated as an essential part of the \ac{sg} infrastructure, encompassing a holistic approach of preventive, detective, and reactive measures~\cite{kawoosa2021review}.
One challenge in integrating countermeasures in the form of active security concepts into power grids is the consideration of legacy compliance, especially for countermeasures that actively interfere with grid operations~\cite{sridhar2011cyber}.
More passive countermeasures such as \ac{ids} offer the opportunity to support cybersecurity observability without imposing restrictive requirements on existing infrastructures, such as the performance specification of endpoint hosts~\cite{krause2021cybersecurity}.
More autonomous \ac{ids} build on \ac{ml} that uses training data from operational or attack scenarios to classify anomalies or false positives from other \ac{ids} to support \ac{soc}~\cite{demertzis2018next}.
However, this requires attack data for the development and validation of these measures, especially data-driven approaches, which often cannot be accessed for security or privacy reasons~\cite{burstein2008toward}.
To address these challenges, attack data are generated synthetically~\cite{zarpelao2020machine} or under laboratory conditions~\cite{ashok2016powercyber}.
Laboratory environments can provide accurate data, but are costly to set up and consequently inflexible and difficult to scale.
Synthetically generated data are more flexible and transferable to other use cases due to their level of abstraction, but can also suffer in quality due to inaccurate modeling and simulation.

To provide a basis for studying the cybersecurity of the \ac{sg} by enabling the generation of attack data, a viable approach that compromises accuracy and effort is required.
More specifically, the challenges we address in this paper are:
\begin{enumerate}[(i)]
    \item Replication of the \ac{sg} with all layers relevant for attack data synthesis and considering multi-stage patterns.
    \item Execution of multi-stage attacks that replicate consistent and flexible attack patterns for dataset diversity.
    \item Generation of adequate data sets using processable formats and consistent data structures based on use cases.
\end{enumerate}

Therefore, in this paper, we propose a multi-stage approach to synthesize cyberattack data that allows generating datasets for \ac{ml} based \ac{ids} in an abstract manner.
To this end, we present an approach based on a knowledge graph that abstracts the network and multi-stage attack procedures in \ac{sg} application scenarios.
In particular, our contributions are:
\begin{enumerate}
	\item We present the state of the art in synthesizing multistage cyberattack data and highlight the challenges of attack data generation (Section~\ref{sec:analysis}).
	\item We describe the overall approach consisting of modeling, knowledge graph representation, and simulation of multi-stage cyberattacks in \ac{sg} (Section~\ref{sec:framework}).
	\item We demonstrate and discuss the dataset generation capabilities and quality of our proposed approach through case studies and comparisons with real datasets (Section~\ref{sec:result}).
\end{enumerate}

\section{Background} \label{sec:background}
In this section, we describe the structure of \acp{sg} (Section~\ref{subsec:background_sg}), anomaly detection (Section~\ref{subsec:background_ad}), multi-stage cyberattacks (Section~\ref{subsec:background_ca}) and attack data synthesis (Section~\ref{subsec:background_adg}).

\subsection{Smart Grid} \label{subsec:background_sg}
Based on \ac{pera} for \ac{ics}, \acp{sg} are composed of five main levels: field devices, control systems, and management/operation as well as corporate systems~\cite{assante2015industrial, hossain2012smart, Zscaler.2022}. %
Field devices, Level 0, make up the physical infrastructure of the \ac{sg} and are responsible for monitoring and controlling various aspects of the \ac{sg}.
Control systems, Level 1, process and analyze the data collected from the field devices to make decisions and perform control actions.
Management systems, Level 2, provide higher-level oversight and coordination of the \ac{sg} and provide a connection to external networks and systems.
The third level, Level 3, handles the management of production processes, such as managing batches, utilizing manufacturing operations management systems and manufacturing execution systems, and maintaining records of data.
Level 4 encompasses systems like enterprise resource planning software, databases, email servers, and other tools that are used for logistics and communication, as well as for data storage.
Lastly, the fifth level, Level 5, is the enterprise network, which is not part of the \ac{ics} but is used to gather data from the \ac{ics} systems to make business decisions.

\subsection{Anomaly Detection} \label{subsec:background_ad}
\acp{sg}, which are a type of \ac{ics}, are vulnerable to cyberattacks.
Therefore, it is important to develop methods for detecting anomalies in network traffic that may indicate the presence of a cyberattack~\cite{baptiste2021systematic}.
One approach to anomaly detection in \acp{sg} is to analyze network traffic patterns~\cite{perales2020madics}.
By analyzing sensor readings, control signals, communication logs, and network traffic it is possible to identify deviations from normal behavior that may indicate the presence of a cyberattack~\cite{wolsing2022ipal}.
Another aspect to consider in anomaly detection for \acp{sg} is the possibility of multi-stage cyberattacks~\cite{sen2022onusing}.
These types of attacks involve multiple steps, each of which may be difficult to detect individually.
Therefore, it is important to consider not only individual anomalies, but also the potential for multiple attacks to be connected in a coordinated fashion.
Overall, anomaly detection in \acp{sg} requires a combination of techniques that can analyze network traffic patterns and identify deviations from normal behavior.
By using \ac{ml} algorithms and other advanced techniques, it is possible to detect anomalies that may indicate the presence of a multi-stage cyberattack and take appropriate action to prevent damage to the system~\cite{kus2022afalse}.

\subsection{Multi-Stage Cyberattack} \label{subsec:background_ca}
Multi-stage cyberattacks on \acp{sg} can be difficult to detect and mitigate due to their complexity and the ability for attackers to adapt and evolve their tactics~\cite{ju2020mckc}.
To better understand and combat these types of attacks, researchers have developed various methods for structuring and analyzing multi-stage cyberattacks in a systematic and structured way.
One common approach is the use of kill-chain like concepts, which divide the attack process into distinct stages such as reconnaissance, weaponization, delivery, and exploitation~\cite{yadav2015technical}.
Another approach is the use of open-source datasets, such as the MITRE ATT\&CK Matrix, which catalogs observed cyberattack incidents and can be used to identify patterns and trends in the attack process~\cite{alexander2020mitre, TheMITRECorporation.1182022}.
The MITRE ATT\&CK Matrix for \acp{ics} is a comprehensive tool used to catalog observed cyberattack incidents in the \ac{ics} environment.
It maps 96 types of attacks to 11 types of tactics and 67 new techniques specific to the \ac{ics} environment.
The matrix is useful resource for generating synthetic multi-stage cyberattack data for \ac{ml}-based anomaly detection and help security professionals understand and detect the various methods and tactics used by adversaries in attacks on \acp{ics}.

\subsection{Attack Data Generation} \label{subsec:background_adg}
The generation of an attack dataset for anomaly detection in \acp{ics} is a critical aspect of security research.
One of the major challenges in generating multi-stage cyberattack data for \ac{ml}-based anomaly detection is the reproducibility of the attack sequence~\cite{choi2019comparison}.
It is essential to design an attack sequence that is suitable for the purpose of the dataset and implement it through automation using an integrated representation.
This allows for the reproduction of the attack sequence and enables the explanation of the abnormal dataset information~\cite{choi2021probabilistic}.
Additionally, generating a diverse range of attack sequences with specific requirements can be difficult, so a model that can reflect more diverse attack sequences is needed.
One approach for generating an attack dataset is to analyze existing attack cases from the viewpoint of the MITRE ATT\&CK framework, which catalogs observed cyberattack incidents and maps them to specific tactics and techniques.
Another approach is to use an integrated representation, such as the hidden Markov model (HMM), to reproduce a sequence of attacks in a systematic and structured way.
This allows for the creation of diverse attack sequences that reflect the reality of observed attacks and meet specific requirements, such as reproducibility and diversity.
\section{Related Work \& Problem Analysis} \label{sec:analysis}
In this section, we present related work (Section~\ref{subsec:analysis_network}) and give a problem statement for this work (Section~\ref{subsec:analysis_problem}).

\subsection{Related Work} \label{subsec:analysis_network}
One direction within this research field are approaches that generate attack data in the lab.
An example of this is the work of Sharafaldin et al.\ ~\cite{sharafaldin2018toward}.
A dataset called CICIDS2017 was generated via a lab experiment.
This contains recent attacks from DoS, DDoS, brute force, XSS, SQL injection, infiltration, port scan, and botnet.
Another interesting approach to generating new datasets is presented by Cordero et al.~\cite{cordero2021generating}.
Cordero et al.\ have developed the ID2T framework, which generates reproducible datasets for \ac{ml} based \ac{ids}.
Instead of generating more datasets, Pandey et al.\ ~\cite{pandey2021gan} attempt to modify existing datasets.
To test this method, the dataset UNSW-NB15 is considered.
This consists of normal traffic and a variety of attacks from nine categories.
With the \ac{dae} and the \ac{wcgan}, additional data are generated and added to the original dataset.
The main goal of this work is to investigate the detection quality of \ac{ids} trained on a synthetic dataset.
For this purpose, we select a suitable \ac{ml} based \ac{ids} method that exploits the structure and temporal dynamics of the dataset.
For instance, Gwon et al.\ ~\cite{gwon2019network} investigate whether \ac{lstm} networks are suitable for binary classification of network packets.
The \ac{lstm} network classifies each network packet as part of normal or as part of malicious traffic.
Oliveira et al.\ ~\cite{oliveira2021intelligent} investigate the performance of different \ac{ml} models for anomaly detection.
In this context, the performance of a Random Forest, a \ac{mlp}, and an \ac{lstm} network on the CIDDS-001 dataset are compared.
The results show that the \ac{lstm} network performs with an accuracy of 99.94\% and a F1 score of 0.91 provides the best results.

\subsection{Problem Analysis} \label{subsec:analysis_problem}
The process of creating datasets in a laboratory environment involves planning and implementing the infrastructure and scenarios, monitoring the system during execution, recording data, and ensuring proper execution.
This process is highly labor-intensive and requires expertise in various subfields of computer science such as networks and cybersecurity.
However, these datasets are not modifiable, making it difficult to add or remove attacks or change network topologies without re-running the experiment.
Additionally, attack data is often underrepresented in these datasets as normal network traffic usually dominates.
Care must be taken to prevent attacks on the network while data is being recorded to avoid compromising the dataset.
This becomes more difficult to ensure depending on the size and complexity of the observed network.
In addition, a separate network must be set up in the laboratory to record normal traffic, or a procedure must be implemented to obscure identifying information in the data for privacy reasons.
There is also the problem of limited diversity in using public datasets to create additional datasets due to their limited availability.
Researchers in \ac{ml} also face the challenge of evaluating the performance of their models using only one dataset, leading to models that may not perform well in real network environments.
This is further compounded by the lack of suitable datasets for training \ac{ml} models.

\section{Abstract Multi-Staged Attack Data Synthesis} \label{sec:framework}
To overcome the problems mentioned in Section~\ref{subsec:analysis_problem}, this paper presents an approach that abstracts multi staged attack data and generates it synthetically.
To achieve this goal, an abstracted data format based on real data must first be devised.
In the second step, a framework for synthetic generation of this data must be developed.
For this work, different forms of data representation for the attack data are examined.
Among them are network packets, network flows and \ac{ids} alert logs.
The decision which of these data representations are most suited for the abstracted data synthesis is based on the following criteria:
\begin{enumerate}
	\myitem{(D1)} Representation of data that is accessible in \ac{ics} and enterprise networks. \label{data_requirements_1}
	\myitem{(D2)} Representation of data that is collected in a standardized format. \label{data_requirements_2}
	\myitem{(D3)} Representation of data that is accessible in a human-readable and understandable form. \label{data_requirements_3}
\end{enumerate}
The investigation of the different forms of representation of the attack data has shown that \ac{ids}-alert logs are suitable for abstraction and synthesis as they fulfill all requirements.
For Requirement~\ref{data_requirements_1}, the survey \textit{A SANS 2021 Survey: \ac{ot}/\ac{ics} Cybersecurity} shows that 62\% of grid operators use signature-based \ac{ids} ~\cite{MarkBristow.2021}.
Requirement~\ref{data_requirements_2} is satisfied as there are standardized file formats for \ac{ids}-alert such as the unified2 format used by the Snort \ac{ids} ~\cite{SnortProject.2020}.
The last Requirement~\ref{data_requirements_3} is met because \ac{ids}-alerts are designed to provide information to grid operators such that they can initiate countermeasures.

\subsection{Requirement Analysis} \label{subsec:requirements_analysis}
In this section, requirements for the data synthesis framework are identified.
The requirements can be divided into functional and non-functional requirements.
Functional requirements describe the desired functionalities of the system while non-functional requirements describe the boundary conditions and quality in which those functionalities must be provided.
The non-functional requirements include:
\begin{enumerate}
	\myitem{(N1)} The framework must not use any additional hardware or virtualization. \label{non_functional_1}
	\myitem{(N2)} The framework must not use additional applications for data generation. \label{non_functional_2}
	\myitem{(N3)} The framework must not use attack signatures from existing \ac{ids}. \label{non_functional_3}
	\myitem{(N4)} The framework must not require existing attack datasets. \label{non_functional_4}
\end{enumerate}
Requirements~\ref{non_functional_1} and~\ref{non_functional_2} ensure that the effort for the configuration and setup does not exceed the effort of lab attempts.
This is necessary because data from laboratory tests are closer to reality and should therefore be preferred if the production effort is the same.
Requirements~\ref{non_functional_3} and~\ref{non_functional_4} ensure that poor data quality of existing datasets or poor documentation of existing \ac{ids} signatures do not affect this work.
The following functional requirements set request to the synthesis framework and the abstract data:
\begin{enumerate}[(i)]
	\myitem{(F1)} Attack data must be generated in the form of identified data representation, i.e.\ as \ac{ids} log files.
	\myitem{(F2)} Attack data must contain \acp{fp}.
	\myitem{(F3)} Attack data must contain \acp{fn}.
	\myitem{(F4)} Networks must be flexible and scalable, i.e.\ must be freely configurable.
	\myitem{(F5)} Networks must be representative, i.e.\ contain Enterprise and \ac{ics} Devices.
	\myitem{(F6)} Attacks must be diverse, i.e.\ be freely configurable.
	\myitem{(F7)} Attacks must be multi-staged.
\end{enumerate}
The framework should generate data in the form of abstracted \ac{ids} alert logs, taking into account the different probabilities of \acp{fn} and \acp{fp} in real-world scenarios.
Users should have the capability to configure networks and attacks for defining flexible and scalable scenarios.
To provide a representative depiction of \ac{sg}, the networks should include devices from both enterprise and \ac{ics} layers.
The framework should be able to simulate both \acp{fn} and \acp{fp} cases based on \ac{ids}-detection accuracy and placement in network.
Multi-stage attacks are more complex and have multiple phases or stages, rather than being a single action.
To simulate different scenarios, the configurable attacks should be able to represent and simulate complex and diverse attack strategies and actions.
This way, the framework will be able to provide a realistic representation of how \ac{ids} behaves.

\subsection{Data Abstraction} \label{subsec:data_abstraction}
In this section, the data format selected for synthesis is reduced to a simplified representation.
In the following, the individual data fields of Snort's unified2 format are used as a basis because it is efficient and can store large amounts of data.
MITRE ATT\&CK matrix is used to abstract the cyberattack data and focus on essential information for cyberattack analysis, such as tactics and techniques used by attackers.
This form of abstraction provides an efficient way to store and analyze large amounts of \ac{ids} data and allows for rapid identification of patterns and trends in the data.
Table~\ref{tab:abstracted_alerts} gives an overview of the data fields of the abstracted data format.
\begin{table}[]
\center
\caption{Data fields of the abstracted ids alerts.}
\begin{tabular}{ll}
\multicolumn{1}{l|}{\textbf{Field}} & \textbf{Description} \\ \hline
\multicolumn{1}{l|}{Source-IP} & IP address of the attacker. \\
\multicolumn{1}{l|}{Target-IP} & IP address of the target. \\
\multicolumn{1}{l|}{Source-Platform} & MITRE platform of the attacker. \\
\multicolumn{1}{l|}{Target-Platform} &  MITRE platform of the target. \\
\multicolumn{1}{l|}{MITRE-Tactic} & Detected MITRE tactic. \\
\multicolumn{1}{l|}{MITRE-Technique} & Detected MITRE technique. \\
\multicolumn{1}{l|}{Sensor-IP} & IP address of the detecting sensor. \\
\multicolumn{1}{l|}{\ac{fp}-Flag} & A flag that indicates if the alert is \ac{fp}. \\
\multicolumn{1}{l|}{Attack-Label} & The attack label indicates which multilevel \\
\multicolumn{1}{l|}{}& attack generated the alert. \\
\end{tabular}
\label{tab:abstracted_alerts}
\end{table}

\subsection{Framework Overview} \label{subsec:framework_overview}
\begin{figure}[h]
\center
\includesvg[scale=0.55]{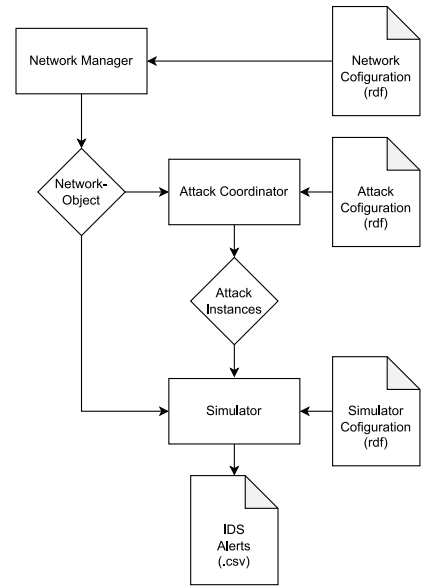}
\caption{Overview of the developed data synthesis framework.}
\label{fig:system_overview}
\end{figure}

The framework architecture is designed to provide an overview of the system for data synthesis.
It is implemented in Python, which is a suitable programming language for prototyping due to its debugging capability and high abstraction.
The system architecture is divided into different modules, as shown in Figure~\ref{fig:system_overview}.
It illustrates the inputs, outputs and interactions of the different modules.
All inputs to the system are in \ac{rdf} data in Turtle format.
For handling \ac{rdf}, the Python library \textit{rdflib}~\cite{RDFLibTeam.rdf.2022} is used, which allows for reading, writing and querying \ac{rdf} files using SPARQL.
To use \ac{rdf} in combination with \ac{shacl}, another library called pyshacl~\cite{RDFLibTeam.pyshacl.2022} is used, which allows for testing \ac{rdf} data against previously defined constraints.
In the following the interaction of the individual modules is addressed.
First of all, the Network Manager reads a network configuration, which contains the description of a network.
Network-Manager then generates a network graph from this information and provides functions for path finding.
Next, the network graph is handed over to the attack coordinator.
This module reads the attack configuration, which defines a multi-staged attack.
Based on this configuration, a set of attack instances is determined by executing the attack on the network graph.
These attack instances, along with the network graph, are then passed to the simulator.
The simulator uses this information to generate \ac{ids} alerts for selected attack instances by checking if the attack communication is detected by an \ac{ids}.
In addition, \ac{fn} and \ac{fp} are generated.
The rate of \ac{fn} and \ac{fp} can be controlled.
Finally, the simulator bundles the generated alerts into a csv file.
Due to the efficient use of the integrated SPARQL query language in rdf, the generation is done in a reasonable time frame.

\subsection{Model Network} \label{subsec:model_network}
The network is modeled by a connected and undirected graph
$G_{net} = (V_{net}, E_{net})$.
In our model, an edge between two nodes $v_1, v_2 \in V_{net}$ exists if and only if there is a defined communication connection between them.
The set of nodes $V_{net}$ consists of four types of nodes, which can be used for constructing a network.
These can be found in Table~\ref{tab:node_types}.
\begin{table}[]
\center
\caption{Types of nodes in the network graph.}
\begin{tabular}{ll}
\multicolumn{1}{l|}{\textbf{Node Type}} & \textbf{Description} \\ \hline
\multicolumn{1}{l|}{Computer} & \parbox{0.3\textwidth}{Computers represent all devices in the network, which provide or use services for the control of the \ac{sg}.} \\
\multicolumn{1}{l|}{Router} & \parbox{0.3\textwidth}{Routers represent the edge router of the network.} \\
\multicolumn{1}{l|}{Switches} & \parbox{0.3\textwidth}{Switches connect the individual components of the network and are possible nodes for the implementation of \ac{ids}.} \\
\multicolumn{1}{l|}{Firewalls} & \parbox{0.3\textwidth}{Firewalls restrict the communication of the network and can be configured by rules.} \\
\end{tabular}
\label{tab:node_types}
\end{table}
Each node must contain at least an IP address and a MITRE platform.
The IP address must be unique, because in this framework it is used to identify nodes.
MITRE platforms are a set of platforms that are referenced as possible targets by MITRE Techniques.
These are later used to find possible targets for the configurable multi-stage attacks.
For both attributes sets can be defined.
The set \textit{IP-addresses} can be defined with a Range of ipv4 or ipv6 addresses and the set \textit{Platforms} is a subset of all MITRE Platforms, which can be selected based on the network being modeled.
For instance, it can include IP addresses in the range of ''172.32.0.0 - 172.47.255.255'', ''192.168.0.0 - 192.169.255.255'' or ''198.20.0.0 - 198.255.255.255''.
\begin{align}
	\text{IP-addresses} =& \{\text{0.0.0.0, ..., 255.255.255.255}\}\\
    \text{Platforms} =& \{\text{Network, Windows, Linux, ..., Containers,}\nonumber \\
    & \text{Control Server, Data Historian, ...,}\}
\end{align}
The Set of all nodes $V_{net}$ consists of the union of all node sets $ V_{net} = \{\text{Computer}\} \cup \{\text{Router}\} \cup \{\text{Switches}\} \cup \{\text{Firewalls}\}$.
In the following the attributes of the individual node types are specified.
Computers and routers contain only ip and platform therefore this can be formally stated as:
\begin{align}
    \text{Computer} =& \{ (p_i, ip_i) | p_i \in \text{Platform}, ip_i \in \text{IP-addresses} \} \\
    \text{Router} =& \{ (p_j, ip_j) | p_j \in \text{Platform}, ip_j \in \text{IP-addresses} \}
\end{align}
Switches must also contain the information whether an \ac{ids} has been implemented.
A Network-based \ac{ids} collects data from a switch via network taps or mirrored ports (SPAN-Ports) configured on the switch to detect and alert on suspicious network activity.
For this an additional boolean variable \textit{IsNIDSActiv} must be added.
If the value is True, it is assumed that a \ac{ids} is active at this switch.
The set of Switches is formally stated as follows:
\begin{align}
\text{Switches} =& \{ (p_k, ip_k, \text{IsNIDSActiv}) | \nonumber \\
&p_k \in \text{Platform}, \nonumber \\
&ip_k \in \text{IP-addresses}, \nonumber \\
&\text{IsNIDSActiv} \in \{True, False\} \}
\end{align}
Finally, the set of firewalls and their rules must be defined.
The rules consist of a tuple with three items, which refers to a switch $s_1$, a switch $s_2$ and a computer $c_1$.
\begin{align}
\text{Firewalls} =& \{(p_l,ip_l,R)| \nonumber\\
&p_l \in \text{Platform},\nonumber\\
&ip_l \in \text{IP-addresses},\nonumber\\
&R \subseteq \text{Rules} \}\\
\text{Rules} =& \{(c_i, s_{k1}, s_{k2})| \nonumber \\
&c_i  \in \text{Computer}, \nonumber \\
&s_{k1}, s_{k2} \in \text{Switches}\}
\end{align}
To understand how the rules work, the possible edges and the concept of communication for $G_{net}$ has to be defined first.
The edges $E_{net}$ of the network graph $G_{net}$ consists of a subset of the cross product of $V_{net}$ with \text{Switches}.
\begin{align}
E_{net} \subseteq \{(e_1, e_2) | e_1 \in V_{net}, e_2 \in \text{Switches}\}
\end{align}
This implies that each edge must be connected to at least one switch.
A device $v_1$ can start to communicate with another device $v_2$ if there exists a path from $v_1$ to $v_2$, where each firewall allows the device $v_1$ to communicate.
A possible paths can contain either no, one, or several firewalls.
In case there is no firewall on the path, device $v_1$ can communicate with device $v_2$.
In the two other cases, each firewall must contain a rule that allows communication.
The evaluation of the firewall rules works the same way in both cases.
Without loss of generality we can consider the following path:
\begin{align}
 v_1 \rightarrow ... \rightarrow s_1 \rightarrow f_1 \rightarrow s_2 \rightarrow ... \rightarrow v_2
\end{align}
For $v_1$ to communicate with $v_2$, the firewall $f_1$ must contain the rule $(s_1,s_2,v_1)$.
Here $s_1$ references the switch that is on the path directly before the firewall, $s_2$ references the switch that is directly after the firewall and $v_1$ is the device which would like to start the communication.
Over the given switches each rule gets a direction, this is reflected in the fact that the communication in the graph $G_{net}$ is not symmetrical.
In this way, the communication in the network can be precisely controlled.\par
\begin{figure}[h]
\center
\includesvg[scale=0.55]{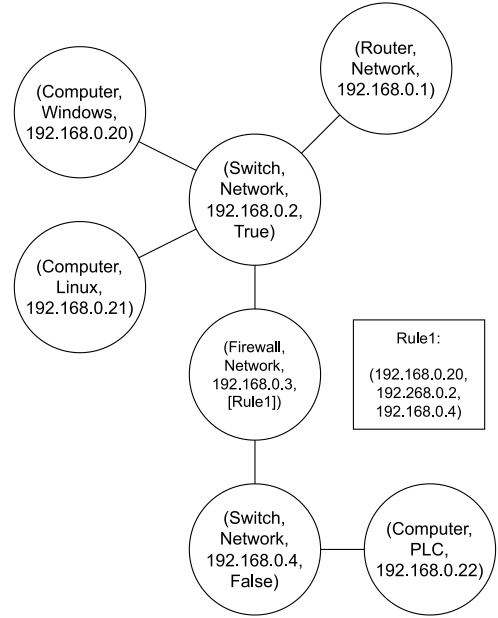}
\caption{An example network graph.}
\label{fig:netzwerk_graph}
\end{figure}

An example of a simple network graph can be seen in Figure~\ref{fig:netzwerk_graph}.
This contains seven nodes, including three computers, two switches, a router and a firewall with a rule.
This rule allows computer with IP 192.168.0.20 to communicate with computer 192.168.0.22.

\subsection{Model Attacks} \label{subsec:model_attacks}
\begin{figure}[h]
\center
\includesvg[scale=0.55]{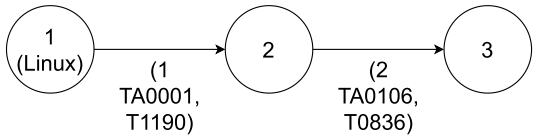}
\caption{A example attack graph.}
\label{fig:attack_graph}
\end{figure}

Multi-staged attacks are represented as directed graphs $G_{att} = (V_{att}, E_{att})$.
It should be noted that these are not attack graphs in a traditional sense.
The set of nodes $V_{att}$ represents wildcards, which later have to be filled with nodes from the network graph to execute an attack.
$V_{att}$ consists of a subset of the natural numbers:
\begin{align}
V_{att} \subseteq & \mathbf{N}
\end{align}
The edge set $E_{att}$ represents the single step attacks that occur between the nodes and together represent the multi-staged attack.
To model the attacks $E_{att}$ the techniques and tactics of the MITRE ATT\&CK Matrix are needed.
Both are represented by a set containing the IDs.
\begin{align}
    \text{Tactics} =& \{\text{T0100, T0101, ..., T0111,} \nonumber \\
        & \text{T0001, T0002, ..., T0011,} \nonumber \\
        & \text{T0040, T0042, T0043}\} \\
    \text{Techniques} =& \{ \text{T0800, T0801, ..., T0891,} \nonumber\\
        & \text{T1001, T1003, ..., T1649}\}
\end{align}
With this information, $E_{att}$ can be formally stated as:
\begin{align}
E_{att} \subset & \{(v_1,v_2, step, tactic, technique)| \nonumber \\
	 & v_1,v_2 \in V, \nonumber \nonumber \\
     & step \in \mathbf{N}, \nonumber \nonumber \\
     & tactic \in \text{Tactics}, \nonumber \nonumber \\
     & technique \in \text{Technique}\}
\end{align}
Each attack configuration also contains exactly one root node from which the attack originates.
It additionally contains a single MITRE platform, which specifies the device type where the attack starts.
For the remaining nodes, possible MITRE platforms can derived from the incoming edges.
For this, the MITRE ATT\&CK matrix is used again.
It provides a mapping $TechniqueToPlatforms$ that assigns each MITRE technique to a subset of the MITRE platforms.
This subset represents possible targets for the technique.
If a node has only one incoming edge, the platforms can be found directly via $TechniqueToPlatforms$.
If a node has several incoming edges, all platforms for the individual techniques must first be determined using $TechniqueToPlatforms$.
After that the intersection of the different platform sets must be formed.
The Platforms that are in the intersection are common goals of the different techniques.
An example of an attack configuration $G_{att}$ can be seen in the Figure~\ref{fig:attack_graph}.

\subsection{Data Synthesis} \label{subsec:data_synthesis}
\begin{figure}[h]
\center
\includesvg[scale=0.55]{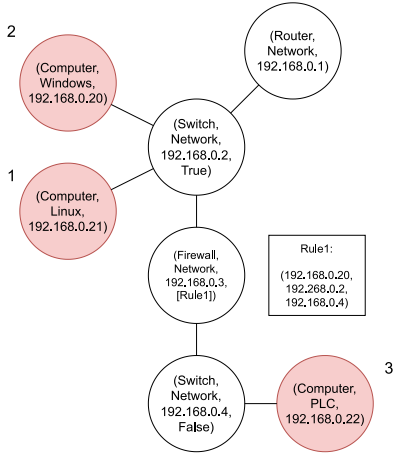}
\caption{A instance of the Attack from Figure~\ref{fig:netzwerk_graph} executed on the network graph from Figure~\ref{fig:attack_graph}.}
\label{fig:netzwerk_attack_graph}
\end{figure}

The attack configurations discussed in Section~\ref{subsec:model_attacks} are now used together with the network configurations from Section~\ref{subsec:model_network} to generate appropriate \ac{ids} alerts.
For this, the instances of the attack configuration for the network configuration must be found.
An instance of an attack is represented by a mapping function, denoted as $F$, which assigns nodes from the set of attack nodes, $V_{att}$, to nodes in the network, $V_{net}$.
For any pair of nodes, $x_1$ and $x_2$, in $V_{att}$ that have an edge between them, denoted as $E x_1,x_2$, the function $F$ must map the nodes such that $F(x_1)$ and $F(x_2)$ are in $V_{net}$ and $F(x_1)$ can communicate with $F(x_2)$, as outlined in Section~\ref{subsec:model_network}.
Additionally, for each node $x_1$ in $V_{att}$, the platform of the node $F(x_1)$ must be present in the platform set of node $x_1$, as outlined in Section~\ref{subsec:model_attacks}.
To find all instances, a complete graph traversal is performed using an algorithm described in Listing~\ref{lst:instance_algorithm}.
The function $network\textunderscore graph.get\textunderscore step\textunderscore values(attack\textunderscore edges)$ is used to find a set of possible solutions for each attack step in $E_{att}$ that comply with the conditions stated above.
These sets are called $step\textunderscore values$.
To generate instances from these partial solutions, the function $solve\textunderscore attack\textunderscore step$ is called for each attack step with the corresponding $step\textunderscore values$.
The result is a list of dictionaries, where each dictionary represents the function $F$ and thus represents an instance of an attack.
\begin{adjustwidth}{6mm}{}
\begin{lstlisting}[
	numbers=left,
	captionpos=b,
	language=Python,
	label=lst:instance_algorithm,
	caption=Algorithm for finding all instances of an attack on a network.,
    basicstyle=\tiny, %
]

def solve_attack_step(attack_edge, step_values, instances):
    new_instances = []
    x1, x2 = attack_edge
    for instance in instances:
        for n1, n2 in step_values:
            i_copy = instance.copy()

            if i_copy[src] is None and i_copy[target] is None:
                i_copy[src] = n1
                i_copy[target] = n2
                new_instances.append(i_copy)

            elif i_copy[src] == n1 and i_copy[target] == n2:
                new_instances.append(i_copy)

            elif i_copy[src] is None:
                if i_copy[target] == n2:
                    i_copy[src] = n1
                    new_instances.append(i_copy)

            elif i_copy[target] is None:
                if i_copy[src] == n1:
                    i_copy[target] = n2
                    new_instances.append(i_copy)
    return new_instances

def find_all_instances(network_graph, attack_graph):
    attack_edges = attack_graph.get_edges()
    step_values = network_graph.get_step_values(attack_edges)
    instances = []
    for i, pair in enumerate(attack_edges):
        instances = solve_attack_step(pair, step_values[i], instances)
    return instances
\end{lstlisting}
\end{adjustwidth}
Figure~\ref{fig:netzwerk_attack_graph} shows an attack instance of the attack defined in Figure~\ref{fig:attack_graph}, executed on the network in Figure~\ref{fig:netzwerk_graph}.
To verify that this instance is valid, there are several conditions that must be met.
Firstly, the root of the attack must be a node that uses the Linux MITRE platform, which is confirmed in this case.
Secondly, the MITRE techniques used in the attack must match the platforms of the target nodes.
In this instance, node 192.168.0.21 uses technique T1190 to attack node 192.168.0.20.
T1190 is known to target Linux, Network, Windows and macOS platforms, thus the platform condition is fulfilled.
In the second step, node 192.168.0.20 uses technique T0836 to attack node 192.168.0.22, T0836 can target Control Server, Field Controller, HM and SIS, fulfilling this condition as well.
Finally, the nodes must be able to communicate with each other.
There is no firewall between nodes 192.168.0.21 and 192.168.0.20, they are connected to a switch, they can communicate with each other.
However, on the path between nodes 192.168.0.20 and 192.168.0.22, there is a firewall which allows communication for node 192.168.0.20 on the path $ 20 \rightarrow 2 \rightarrow 3 \rightarrow 4 \rightarrow 22$ (IP addresses without Prefix 192.168.0.).
Therefore, all conditions are met and this instance is valid.
The next step is to convert the instances of the attack into \ac{ids} alerts.
This process requires the mapping function $F$, and the network graph $G_{net}$ and the attack graph $G_{att}$.
For each attack edge $(x_1,x_2) \in E_{att}$, it is verified if the path between the nodes $F(x_1)$ and $F(x_2)$ in the network graph contains a switch that is configured as \ac{ids}.
If such a switch exists, an alert is generated with the attributes listed in Table~\ref{tab:abstracted_alerts}.
The IP addresses and MITRE platforms are obtained from the network graph $G_{net}$, while the MITRE tactic, technique and attack label are obtained from the corresponding edges in the attack graph $G_{att}$.
These alerts are considered \acp{tp}, thus the \ac{fp} flag is set to False.
The alerts generated for the instance in Figure~\ref{fig:netzwerk_attack_graph} can be found in Table~\ref{tab:generierte_alerts}.
\begin{table}
	\center
    \caption{This table shows the alerts that were generated with the example instance in Figure~\ref{fig:netzwerk_attack_graph}.}
	\begin{tabular}{lll}
        \textbf{Name} & \textbf{Alert 1} & \textbf{Alert 2}\\
        \hline
        Source-IP & 192.168.0.21 & 192.168.0.20 \\
        Target-IP & 192.168.0.20 & 192.168.0.22 \\
        Source-Plattform & Computer & Computer \\
        Target-Plattform & Computer & Field Controller\\
        MITRE-Taktik & TA0001 & TA1190 \\
        MITRE-Technik & TA0106 & TA0106 \\
        Sensor-IP & 192.168.0.2 & 192.168.0.2 \\
        \ac{fp}-FLAG & False & False \\
        Attack-Label & Attack\_1 & Attack\_1
    \end{tabular}
    \label{tab:generierte_alerts}
\end{table}
There are two methods for adding \acp{fn} to the generated \ac{ids} alerts.
One method is to selectively omit certain \ac{ids} devices from the network, which will increase the rate of \acp{fn} since attacks will go undetected.
Another method is to randomly remove \acp{tp} from the data, which allows for more precise control over the proportion of \acp{fn}.
This second method will be used here.
In addition to generating \acp{fn} and \acp{tp}, it is also necessary to generate \acp{fp}.
To do so, two nodes, $x_1,x_2$, are randomly selected from the network graph $G_{net}$, and it is checked if $x_1$ can communicate with $x_2$, and if there is an \ac{ids} device on the communication path.
If such nodes are found, a suitable attack is randomly selected from the MITRE ATT\&CK matrix and an alert is generated with it.

\section{Evaluation \& Discussion} \label{sec:result}
In this section, we evaluate the effectiveness of using synthesized data for \ac{ml}-based \ac{ids} by testing the performance of \ac{mlp} and \ac{lstm} models, which are trained with synthesized data.
The models are tested on both synthesized data and a data set from a laboratory experiment, evaluating both alert classification and attack detection.
Additionally, the models are tested with both the full set of alert attributes and a reduced set of attributes that exclude the MITRE Tactic and Technique.

\subsection{Procedure for Investigation} \label{subsec:result_procedure}
The investigation will be carried out as follows.
First, configurations for the data synthesis framework are created for various scenarios.
Then \ac{ids}-alert data sets with the required properties (\ac{tp}-rate \ac{fn}-rate and \ac{fp}-rate) are generated.
Finally, \ac{mlp} and \ac{lstm} models are trained on this data and the performance will be measured.
For the \ac{lstm} network, 2 layers are used.
The first is an \ac{lstm} layer with 300 units and a dropout of 0.2 units and a dropout of 0.2.
This layer receives a two-dimensional input whose size is equal to the feature vector times the length of the sequence.
Three layers are used for the \ac{mlp}.
These include an input layer, a hidden layer, and an output layer.
The input layer has the size of the feature vector.
For the hidden layer, a dense layer is always used in the \ac{mlp}.
This has a size of 200 and uses the ReLU function as activation function.
To give an indication of the performance on real data, the \ac{ml} models, which are trained on the synthesized data, getting tested with the data from a lab experiment.
For this purpose, the \ac{pcap} file of the laboratory experiment must be manually transferred to the abstract data format (cf.\ Section~\ref{subsec:data_abstraction}).

\subsection{Scenarios} \label{subsec:scenarios}
The network configuration used in the study were developed using the \ac{pera} and the \ac{nist} publication \textit{Guide to Industrial Control Systems Security}.
The \ac{nist} publication contains recommendations for securing \ac{ics} networks and, in particular, structures for network architectures ~\cite{Stouffer.2015}.
The multi-stage attack configurations used in the study were created using the \ac{ics} kill chain and the MITRE software database.
The \ac{ics} kill chain describes cyberattacks that aim to destroy the \ac{ics} infrastructure or temporarily disrupt production.
The MITRE software database divides software into tools and malware.
For each software, the MITRE techniques used are listed.
In addition, each technique has an explanation of exactly what it is used for.
The created configurations replicate attacks of known malware.
Among them are Stuxnet, Industroyer, WannaCry and PLC-Blaster.

\subsection{Data Set Creation} \label{subsec:dataset_creation}
To generate data with the presented framework in a way that \ac{ml} models can utilize them, larger datasets have to be generated.
In this work, datasets were generated by producing \ac{ids} alerts for a specific network configuration with a set of different attacks.
The number of alerts for an attack configuration can be chosen arbitrarily.
In addition, attributes of the whole dataset, such as the ratio of \ac{tp} to \ac{fp} or \ac{tp} to \ac{fn}, can be controlled.
This is helpful to generate the appropriate training datasets for the different applications of the \ac{ml} models.
In our case these are the alert classification and the attack detection.

\subsection{PCAP Transformation} \label{subsec:pcap_transformation}
The data of the laboratory experiment contains the network traffic of a test network in which the WannaCry ransomware is spreading~\cite{DavidSzili.2017}.
To make the network packets of a \ac{pcap} file usable for our \ac{ml} models, they first have to be converted into the abstracted \ac{ids} alerts.
To generate \ac{ids} alerts from a \ac{pcap} file, the \ac{ids} Snort is used.
Snort is a rule-based \ac{ids} and can also apply rules to archived data in the form of \acp{pcap} ~\cite{SnortProject.2020}.
In this work the Emerging Threats ruleset is used ~\cite{ProofpointInc.2022}.
By applying the ruleset, the \ac{pcap} file with 46654 network packets is converted into a log file with 7874 alerts.
This list contains a lot of duplicate alerts.
To filter them out, duplicates that follow each other directly in time are removed.
This reduces the number of alerts to 5406.
These generated alerts contain a timestamp, the rule ID, a description, the protocol and the IP addresses.
The number of alerts must be further reduced to manually assign labels.
For this purpose, the priorities specified in the definition of each rule are used.
All alerts with a low priority are removed.
After this step, 66 alerts remain, which were generated by seven different rules.
For these 66 alerts we have to manually assign MITRE ATT\&CK tactics and techniques, while the alerts generated by the same rule get the same tactics or techniques.
Using the manually assigned MITRE information and the knowledge about the platforms of the devices in the \ac{pcap}, the 66 alerts are transferred into the abstracted alert format.
The Sensor ID attribute is ignored because the \ac{pcap} file does not contain any information about sensors.
The \ac{fp} label is set to False and the Attack label is set to WannaCry, since the \ac{pcap} contains only the WannaCry attack.
\acp{fp} are needed to complete the dataset.
These are generated using the synthesis environment for the period of the WannaCry attack.
The final data set has 10 times as many \ac{fp} as \ac{tp}.
Thus, the data set consists of 66 \ac{tp} and 660 \acp{fp}.

\subsection{Result} \label{subsec:result}
\begin{figure}[h]
\center
\includesvg[scale=0.60]{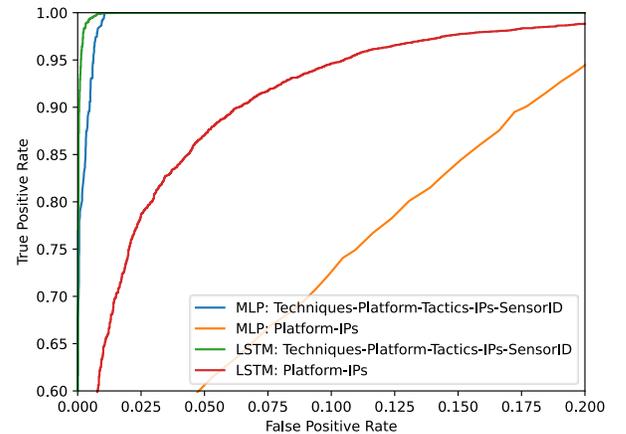}
\caption{Alert classification results on synthesized data. $x_{axis} \in \{0, 0.2\}$ and $y_{axis} \in \{0.6, 1.0\}$}
\label{fig:roc-alert-classification-synthetic-data}
\end{figure}

\begin{figure}[h]
\center
\includesvg[scale=0.60]{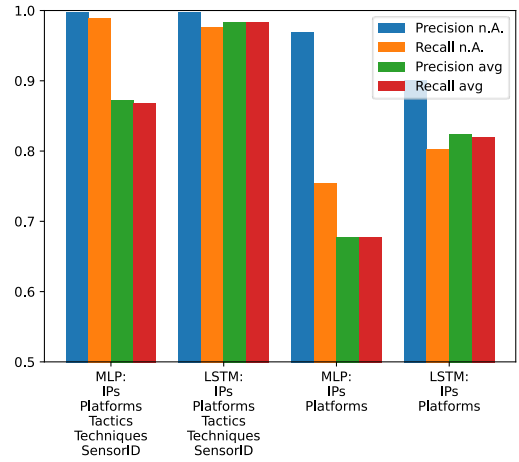}
\caption{Attack detection results on synthesized data (n.A = no Attack). $y_{axis} \in \{0.5, 1.0\}$}
\label{fig:precision-recall-attack-detection-synthetic-data}
\end{figure}

\begin{figure}[h]
\center
\includesvg[scale=0.60]{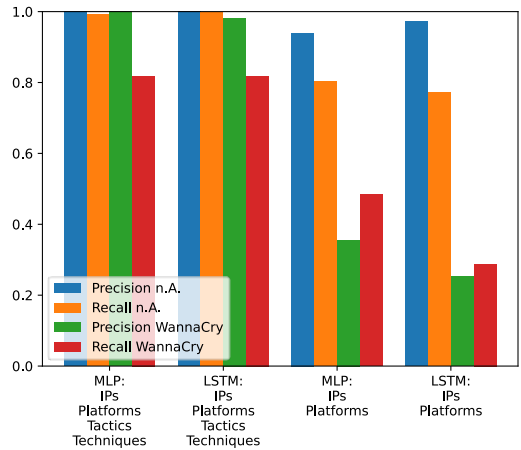}
\caption{Attack detection results on laboratory data (n.A = no Attack).}
\label{fig:precision-recall-attack-detection-labor-data}
\end{figure}

The results in this section were generated through a thorough process of training and testing the \ac{ml} algorithms multiple times.
The consistency in the results indicates the robustness of the models and the lack of significant difference in multiple runs of the algorithm makes it unnecessary to include standard deviation in the graphs.
Additionally, the models have not been fine-tuned to their full potential, indicating room for improvement.
Therefore, these results can be considered as a strong starting point for further optimization and refinement.
The first experiment tests the performance of the \ac{ml} models for alert classification.
Here, the \ac{ml} models should recognize the \ac{tp} and thus separate them from the \ac{fp} to simplify the manual processing of \ac{ids} alerts by the operator of the \ac{sg}.
For this two datasets were generated with $0\%$ \ac{fn} and with a \ac{tp} to \ac{fp} ratio of $100\%$.
Alerts for attacks were generated until 2000 alerts were collected for each attack mentioned in Section~\ref{subsec:scenarios}.
This means that the datasets contain a total of 8000 \ac{tp} and 8000 \ac{fp}, this results in a balanced dataset.
The \ac{mlp} and \ac{lstm} are trained on the first data set and tested on the second one.
The results of this test can be shown as a roc curve because we are looking at a binary classification tested on a balanced dataset.
This roc curve is shown in Figure~\ref{fig:roc-alert-classification-synthetic-data}.
It was expected that the \ac{lstm} would perform better than the \ac{mlp}, since the \ac{lstm} can correlate successive alerts and thus derive further insights.
The \ac{mlp}, on the other hand, can only look at individual alerts.
The roc curve shows that the \ac{lstm} performs better on the whole data set as well as on the data set which only uses the attributes source-IP, target-IP, source-platform and target-platform.
Next, the performance of the \ac{ml} models trained on the synthesized data is tested on the laboratory data set from Section~\ref{subsec:pcap_transformation}.
The results are visualized here as bar diagram of the \ac{tp} rate and the \ac{fn} rate.
The roc curve cannot be used here because this is not a balanced data set.
On the complete dataset the performance of the models is very good, reaching a \ac{tp} rate of $0.99$ and on the subset of the data set a \ac{tp} rate over $0.8$.
The expected performance difference between the two models cannot be confirmed here.
Here, the \ac{lstm} model no longer provides the expected advantage.
\begin{figure}[h]
\center
\includesvg[scale=0.60]{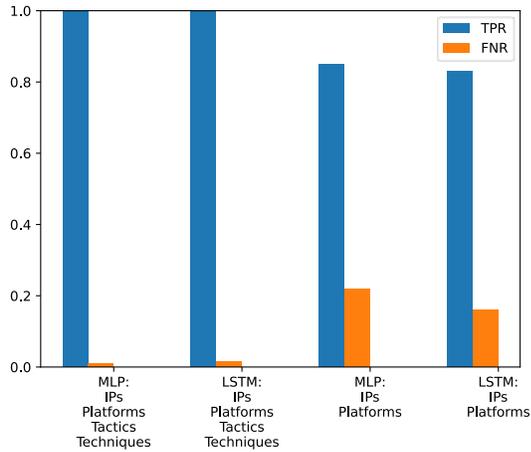}
\caption{Alert classification results on laboratory data.}
\label{fig:tpr-fnr-alert-classification-labor-data}
\end{figure}
The next experiment tests the performance of the \ac{ml} models for attack detection.
Here, the \ac{ml} models should recognize the ongoing Attack so that \ac{sg} operators can take appropriate countermeasures.
For this two data sets were generated with $0\%$ \ac{fn} and with a \ac{tp} to \ac{fp} ratio of $0.25$.
Again, alerts were generated until 2000 alerts were collected for each attack.
This means that the data sets contain a total of 2000 for each attack and 2000 \ac{fp}, so again balanced but over 5 classes.
In this experiment, the \ac{ml} models must assign the alerts to one of these 5 classes: Stuxnet, Industroyer, WannaCry and PLC-Blaster or False-Positive.
The \ac{mlp} and \ac{lstm} are trained on the first data set and tested on the second one.
The results of this test can be shown as bar diagram, which takes the accuracy and recall into account.
The class of false positives (n.A. = no attack) is considered individually, since there should be as few \ac{tp} alerts as possible sorted into this class to not miss any attacks in the network.
The results are shown in Figure~\ref{fig:precision-recall-attack-detection-synthetic-data}.
The results for the detection of \ac{fp} (or no attack) for both models do not differ significantly, just as in the previous experiment.
This can be seen from the fact that the Precision and the Recall for the class \textit{no Attack} are in the same region.
When detecting to which attack an alert belongs, alerts must be correlated to improve detection performance, since the same alert can belong to multiple attacks.
This reveals the strength of \ac{lstm}.
The average precision and recall across all classes is 10 percentage points higher with \ac{lstm} than with \ac{mlp}.
This holds for the whole data set as well as for the subset.
Lastly, the performance of the \ac{ml} models trained on synthesized data of the previous experiment is tested on the laboratory dataset.
The results are shown in Figure~\ref{fig:precision-recall-attack-detection-labor-data}.
Again, the advantages of the \ac{lstm} are no longer apparent.

\subsection{Discussion} \label{subsec:result_dis}
In this section, the results of Section~\ref{subsec:result} are critically reviewed.
Although the results of the \ac{ml} models are promising, this is not an indication that proposed approach is transferable to real scenarios for the following reasons.
The results of the \ac{lstm} were always better than those of the \ac{mlp} in the experiments with the synthetic data.
This indicates that the synthetic data contains a structure that can be recognized by the \ac{lstm}.
The data from the laboratory experiment did not show such a clear performance advantage for the \ac{lstm}.
This suggests that the attacks in the lab data do not have the same structure.
This could be due to the manual steps necessary to convert the \ac{pcap} into a usable format.
A second reason could be that in the data set the \ac{fp} is so strongly overweighted.
This could have made the examined sequence length of the \ac{lstm} too small.
Another point that needs to be addressed is the performance of the \ac{ml} models.
The \ac{tp} rate of the models is very close to 100\%.
This is usually a sign that the models are being used incorrectly or that the training data contains data that should not be there, such as labels.
In our case, it is likely that the models can separate the \ac{tp} and \ac{fp} alerts too well via the MITRE tactic and technique attribute.
In the attack configurations only a small subset of all techniques are used.
This means that a large part of the \acp{fp} can be detected by their techniques, because they are exclusively used in the \acp{fp}.
This leads to the next problem is the \ac{fp} generation.
The \ac{fp} from the synthesis environment are randomly generated.
As already stated this is a problem because the attacks configured in the study does not cover all techniques.
\ac{fp} found in real environments are not random and can occur for many different reasons, such as bad configurations or hardware errors.
Also, the framework and the \ac{ml} models were not tested extensively.
The reason for this is that all configurations in this work were configured manually.
The results were generated with a set of 4 attack configurations and only one network configuration.
To make a statement about whether the synthesized data can be used for \ac{ml} based \ac{ids}, these test cases should be greatly expanded.
Nevertheless, synthetic data has been shown to have many advantages.
Among them are availability and the flexibility to generate data for different problems.
Another advantage is that the data can be generated in a way that they are optimally suited for training \ac{ml} models.

\section{Conclusion} \label{sec:conclusion}
To improve the security of \acp{sg}, \ac{ml}-based \acp{ids} are being implemented to detect attacks at an early stage.
However, these systems require training data, which is often not available or suitable.
This work investigated whether abstracted and synthetically generated data can be used to train \ac{ml}-based \acp{ids}.
The investigation focused on \ac{ids} log files as a suitable data format for abstraction and synthesis.
A system was presented that simulates multi-stage attacks on a network graph using the MITRE ATT\&CK matrix, and generates \ac{ids} log files containing \acp{tp}, \acp{fp}, and \acp{fn}.
The influence of the individual parameters of the abstracted data format on the performance of the \ac{ml} models was investigated, and it was found that the MITRE ATT\&CK related information in the \ac{ids} log files had the largest influence on their performance.
The \ac{ml} models were also tested on a dataset containing real attack data and were able to separate \ac{tp} from \ac{fp} and assign them to correct attacks.
The resulting performance is promising in terms of training anomaly detection systems for SG without requiring a large number of datasets for training.
Future work will address the study of synthetic, multi-stage cyberattacks in laboratory environments using digital twin approaches where the reference grid is replicated in our framework and a comparision with the lab is performed.
In addition, other constellations of dataset construction will be investigated, e.g.\ the degree of blending of real and synthetic data also in terms of recognition and classification performance.


\end{document}